\documentclass[a4paper,preprintnumbers,amsmath,amssymb,twocolumn,10pt]{revtex4-2}

\usepackage{graphicx}% Include figure files
\usepackage{dcolumn}% Align table columns on decimal point
\usepackage{bm}% bold math
\usepackage{epstopdf}
\newcount\driver
\newcount\bozza

scaled\magstep1                  
scaled\magstep1 
 
scaled\magstep1                 \font\indbf=cmbx10 scaled\magstep2

scaled \magstep2

{\count255=\time\divide\count255 by 60
\xdef\hourmin{\number\count255}
        \multiply\count255 by-60\advance\count255 by\time
   \xdef\hourmin{\hourmin:\ifnum\count255<10 0\fi\the\count255}}

\let\a=\alpha     \let\g=\gamma     \let\d=\delta     \let\e=\varepsilon
  \let\h=\eta           \let\l=\lambda
\let\m=\mu    \let\n=\nu      \let\x=\xi                \let\r=\rho
\let\s=\sigma \let\t=\tau            
\let\ps=\psi        
\let\G=\Gamma        \let\L=\Lambda

\def\EE{{\cal E}}\def\VV{{\cal V}}

\def\TT{{\cal T}}

\def\nn{{\bf n}}

       \def\oo{{\underline \omega}}
\def\ee{{\underline \varepsilon}}

        \def\EE{\hbox{\msytw E}}

\let\==\equiv

\let\io=\infty
\let\0=\noindent

\def\*{{\hfill\break\null\hfill\break}}

\def\tilde#1{{\widetilde #1}}

\def\la{{\langle}}
\def\ra{{\rangle}}

\def\tende#1{\,\vtop{\ialign{##\crcr\rightarrowfill\crcr
             \noalign{\kern-1pt\nointerlineskip}
             \hskip3.pt${\scriptstyle #1}$\hskip3.pt\crcr}}\,}
\def\otto{\,{\kern-1.truept\leftarrow\kern-5.truept\to\kern-1.truept}\,}

\def\wh#1{\widehat{#1}}
\def\hat#1{\wh{#1}}
\def\sqt[#1]#2{\root #1\of {#2}}

\def\bp{{\bar \ps}}

\def\EE{{\cal E}}\def\VV{{\cal V}}

\def\TT{{\cal T}}

\def\T#1{{#1_{\kern-3pt\lower7pt\hbox{$\widetilde{}$}}\kern3pt}}
\def\VVV#1{{\underline #1}_{\kern-3pt
\lower7pt\hbox{$\widetilde{}$}}\kern3pt\,}
\def\W#1{#1_{\kern-3pt\lower7.5pt\hbox{$\widetilde{}$}}\kern2pt\,}

\def\indica{\leaders \hbox to 0.5cm{\hss.\hss}\hfill}
\def\guida{\leaders\hbox to 1em{\hss.\hss}\hfill}
\mathchardef\oo= "0521

\def\nn{{\bf n}}

\def\oo{{\underline \omega}}

\def\qed{\raise1pt\hbox{\vrule height5pt width5pt depth0pt}}
 
  \def\bp{{\bar p}} 

\def\indic{\hbox{\raise-2pt \hbox{\indbf 1}}}

\def\ins#1#2#3{\vbox to0pt{\kern-#2 \hbox{\kern#1 #3}\vss}\nointerlineskip}

\newdimen\xshift \newdimen\xwidth \newdimen\yshift
\def\insertplot#1#2#3#4#5#6{%
\xwidth=#1pt \xshift=\hsize \advance\xshift by-\xwidth \divide\xshift by 2%
\begin{figure}[ht]
\vspace{#2pt} \hspace{\xshift}
\begin{minipage}{#1pt}
#3 \ifnum\driver=1 \griglia=#6
\ifnum\griglia=1 \openout13=griglia.ps \write13{gsave .2
setlinewidth} \write13{0 10 #1 {dup 0 moveto #2 lineto } for}
\write13{0 10 #2 {dup 0 exch moveto #1 exch lineto } for}
\write13{stroke} \write13{.5 setlinewidth} \write13{0 50 #1 {dup 0
moveto #2 lineto } for} \write13{0 50 #2 {dup 0 exch moveto #1
exch lineto } for} \write13{stroke grestore} \closeout13
\includegraphics{griglia.ps} \fi
\includegraphics{#4.ps}\fi%
\ifnum\driver=2 \fi
\end{minipage}
\caption{#5}
\end{figure}
}
\newdimen\shift \shift=-1.5truecm
\def\lb#1{%
\ifnum\bozza=1
\label{#1}\rlap{\hbox{\hskip\shift$\scriptstyle#1$}}
\else\label{#1} \fi}

\def\be{\begin{equation}}
\def\ee{\end{equation}}
\def\bea{\begin{eqnarray}}\def\eea{\end{eqnarray}}
\def\bean{\begin{eqnarray*}}\def\eean{\end{eqnarray*}}
\def\bfr{\begin{flushright}}\def\efr{\end{flushright}}
\def\bc{\begin{center}}\def\ec{\end{center}}
\def\bal{\begin{align}}\def\eal{\end{align}}
\def\ba#1{\begin{array}{#1}} \def\ea{\end{array}}
\def\bd{\begin{description}}\def\ed{\end{description}}
\def\bv{\begin{verbatim}}\def\ev{\end{verbatim}}
\def\nn{\nonumber}
\def\Halmos{\hfill\vrule height10pt width4pt depth2pt \par\hbox to \hsize{}}
\def\pref#1{(\ref{#1})}

\def\ins#1#2#3{\vbox to0pt{\kern-#2 \hbox{\kern#1 #3}\vss}\nointerlineskip}

\newdimen\xshift \newdimen\xwidth \newdimen\yshift
\newcount\griglia

\def\insertplot#1#2#3#4#5#6{%
\xwidth=#1pt \xshift=\hsize \advance\xshift by-\xwidth \divide\xshift by 2%
\begin{figure}[ht]
\vspace{#2pt} \hspace{\xshift}
\begin{minipage}{#1pt}
#3 \ifnum\driver=1 \griglia=#6
\ifnum\griglia=1 \openout13=griglia.ps \write13{gsave .2
setlinewidth} \write13{0 10 #1 {dup 0 moveto #2 lineto } for}
\write13{0 10 #2 {dup 0 exch moveto #1 exch lineto } for}
\write13{stroke} \write13{.5 setlinewidth} \write13{0 50 #1 {dup 0
moveto #2 lineto } for} \write13{0 50 #2 {dup 0 exch moveto #1
exch lineto } for} \write13{stroke grestore} \closeout13
\includegraphics{griglia.ps} \fi
\includegraphics{#4.ps}\fi%
\ifnum\driver=2 \fi
\end{minipage}
\caption{#5}
\end{figure}
}
\newdimen\shift \shift=-1.5truecm
\def\lb#1{%
\label{#1}\rlap{\hbox{\hskip\shift$\scriptstyle#1$}}
\else\label{#1} \fi}

\def\be{\begin{equation}}
\def\ee{\end{equation}}
\def\bea{\begin{eqnarray}}\def\eea{\end{eqnarray}}
\def\bean{\begin{eqnarray*}}\def\eean{\end{eqnarray*}}
\def\bfr{\begin{flushright}}\def\efr{\end{flushright}}
\def\bc{\begin{center}}\def\ec{\end{center}}
\def\bal{\begin{align}}\def\eal{\end{align}}
\def\ba#1{\begin{array}{#1}} \def\ea{\end{array}}
\def\bd{\begin{description}}\def\ed{\end{description}}
\def\bv{\begin{verbatim}}\def\ev{\end{verbatim}}
\def\nn{\nonumber}
\def\Halmos{\hfill\vrule height10pt width4pt depth2pt \par\hbox to \hsize{}}
\def\pref#1{(\ref{#1})}

\usepackage{amsmath}
\usepackage{amsfonts}
\usepackage{amssymb}
\usepackage{epstopdf}

scaled\magstep1

\let\a=\alpha   \let\g=\gamma  \let\d=\delta
\let\e=\varepsilon
  \let\h=\eta     \let\l=\lambda
\let\m=\mu    \let\n=\nu    \let\x=\xi         \let\r=\rho
\let\s=\sigma \let\t=\tau    
\let\ps=\Psi   
\let\G=\Gamma   \let\L=\Lambda

\def\EE{{\cal E}} \def\VV{{\cal V}}
 
\def\TT{{\cal T}}

\def\nn{\nonumber}

\def\\{\hfill\break}
\def\={:=}
\let\io=\infty
\let\0=\noindent

\def\tende#1{\,\vtop{\ialign{##\crcr\rightarrowfill\crcr\noalign{\kern-1pt
    \nointerlineskip} \hskip3.pt${\scriptstyle #1}$\hskip3.pt\crcr}}\,}
\def\otto{\,{\kern-1.truept\leftarrow\kern-5.truept\to\kern-1.truept}\,}

\def\wh{\widehat}
\def\to{\rightarrow}
\def\la{\left\langle}
\def\ra{\right\rangle}
\def\qed{\hfill\raise1pt\hbox{\vrule height5pt width5pt depth0pt}}

\def\be{\begin{equation}}
\def\ee{\end{equation}}
\def\bp{\begin{pmatrix}}
\def\ep{\end{pmatrix}}
\def\bea{\begin{eqnarray}}
\def\eea{\end{eqnarray}}
\def\nn{\nonumber}
\def\pref#1{(\ref{#1})}

\def\lb{\label}

\begin{document}

\title{Nonperturbative renormalization of the lattice Sommerfield vector model}
 
\author{Vieri Mastropietro}

\address{Institute for Advanced Study, Princeton, USA}
\address{University of Milan, Italy}

\begin{abstract} The lattice 
Sommerfield model, describing 
a massive vector gauge field coupled to a light fermion in 2d, is an ideal candidate 
to verify perturbative conclusions. In contrast with continuum exact solutions, we prove that there is no infinite field renormalization, 
implying
the reduction of the degree of the ultraviolet divergence,
and that the anomalies are non renormalized.
Such features are the counterpart of analogue properties at the basis of the Standard Model perturbative renormalizability. 
The results are non-perturbative, in the sense that 
the averages of invariant observables are expressed in terms of convergent expansions
uniformly in the lattice and volume.
%We consider a lattice version of the Sommerfield model, describing a massive vector gauge field coupled to a light fermion in 2d. We prove that
%the correlations are expressed by series expansions which are convergent
%uniformly in the lattice and with a finite field renormalization.
%The analogue of properties at the basis of perturbative renormalizability
%of 4d gauge theory, like the 
%anomaly non-renormalization and the reduction of the degree of divergence by current conservation, are rigorously proven to be valid in this case at a non-%perturbative level.
\end{abstract} 

\maketitle

\section{Introduction.}

Most properties of the Standard Model are known only at a perturbative level with
series expansions expected to be generically diverging; in particular its 
perturbative renormalizability \cite{W},\cite{TH} relies on two crucial properties, the reduction of the degree of divergence with respect to power counting and the cancellation of the anomalies \cite{B111} ensured by the Adler-Bardeen theorem
\cite{AB}. 
Such properties are essential to maintain the renormalizability present
with massless bosons.
The phenomenon of the reduction of the degree of divergence can be already seen in 
a $U(1)$ gauge theory like QED.
Adding a mass to the photon breaks gauge invariance
and produces a propagator of the form ${1\over k^2+M^2}(\d_{\m\n}+ {k_\m k_\n\over M^2})$; due to the lack of decay
of the second term the theory becomes dimensionally non renormalizable. However 
the transition in a $U(1)$ gauge theory like QED from a $M=0$ to a $M\not=0$ case 
is soft and 
the theory remains perturbatively renormalizable \cite{O}; the photons are coupled to a conserved current
$k_\m \hat j_\m=0$ so that the contribution of the non-decaying part of the propagator is vanishing.
A similar reduction happens in the electroweak sector, 
but the fermion mass 
violates the chiral symmetry and 
leads to the Higgs introduction; again the renormalizability proof relies on the fact that the $k_\m k_\n$
term in the propagator does not contribute  \cite{TH}. The chiral symmetry is generically violated
by anomalies which need to cancel out, and such cancellation is based on the Adler-Bardeen property.

All the above arguments are valid in perturbation theory and non-perturbative effects could be missed. 
This issue would be solved
by a non-perturbative lattice anomaly-free formulation of gauge theory, which is still out of reach, see for instance \cite{8}-\cite{12}. In particular one needs to get
high values of cut-off, exponential in the inverse coupling, a property which is the non-perturbative analogue of renormalizabiliy. The implementation of the Adler-Bardeen
theorem and of the reduction of the degree of divergence in a non-perturbative context is however a non trivial issue, as their perturbative  derivation uses dimensional regularizations,  and functional integral derivations
\cite{Fu} are essentially one loop results \cite{AB1}. 

It is convenient therefore to investigate such properties 
in a simpler context, and the Sommerfield model \cite{GRa},
describing a massive vector gauge field coupled to a light fermion in 2d,
appears to be the ideal candidate, see also  \cite{GRb},\cite{GRc},\cite{GRd}.
More exactly, we consider a version of this model with non zero fermionic mass, but our results are uniform in the mass.
The model can be seen as a $d=2$ QED with a massive photon; as in 4d, at the level of perturbation theory 
the transition from $M=0$ to a $M\not=0$ is soft and 
the theory remains super-renormalizable. Again this follows
from the conservation of currents, which is ensured at the level
of correlations by dimensional regularization;
the same regularization provides
the anomaly non-renormalization \cite{GRb}. 
In this case however we have access to non-perturbative information and we can check such conclusions.
Exact solutions are known in the continuum version of the Sommerfield model
\cite{B1}, \cite{GRa}, \cite{H}, 
\cite{D1}. Remarkably the above perturbative features are {\it not} verified; there is 
an infinite wave function renormalization incompatible with the superrenormalizability, and anomalies have a value depending on the regularization.

In this paper we consider the Sommerfield model on the lattice, and we analyze it
using the methods of constructive renormalization. The lattice preserves a number of symmetries, in the form of Ward Identities.
Our main result is that there is no infinite field renormalization, 
which is the counterpart of superrenormalizability,
and that the Adler-Bardeen theorem holds with finite lattice. 
Non perturbative violation of the above perturbative conclusions is therefore excluded. 
Other 2d models previously rigorously constructed, see \cite{B}-\cite{D111},
lack of these features. 
Quantum simulations of 2d models \cite{Ri1}-\cite{Ri3} have been also considered in the literature,
but they regard mostly
the Schwinger model, to which the Sommerfield model reduces when the boson and fermion mass is vanishing.
Our results are non-perturbative, in the sense that 
the averages of gauge invariant observables are expressed in terms of convergent expansions
uniformly in the lattice and volume.

The paper is organized in the following way. In \S II we define a lattice version 
of the Sommerfield model. In \S III we derive exact Ward Identities for the model. In \S IV we integrate the boson field and in \S V we perform a nn-perturbative multiscale analysis for the fermionic fields. In \S VI we prove the validity of the Adler-Bardeen theorem and in \S VII the conclusions are presented. 

\section{The lattice Sommerfield model} 

If $\g_0=\s_1$, $\g_1=\s_2$, we define 
\be
\la O\ra= {1\over Z}
\int \prod_x d\bar\psi_x d\psi_x \int_{R^{2|\L|} } 
\prod_{x,\m} dA_{\m,x} e^{-S(A,\psi)} O
\label{llll}
\ee
where $Z$ the normalization,
$x\in \L$, with $\L$ is a square lattice with step $a$ with antiperiodic boundary conditions and 
\be
S(A,\psi)=S_A(A)+S_\psi(A,\psi)
\ee
with
\bea
&&S_A(A)=a^2 \sum_{x}[{1\over 4}F_{\m,\n,x} F_{\m,\n,x}+{M^2\over 2}
 A_{\m,x} A_{\m,x}]\nn\\
&&S_\psi(A,\psi)=a^2\sum_x [
\tilde m  \bar\psi_x \psi_x+\\
&&a^{-1} Z_\psi
(\bar\psi_x \g^+_\m e^{i a e A_{\m,x}}
\psi_{x+a_\m} -\bar\psi_{x+a_\m} \g^-_\m e^{-i a e A_{\m,x}}
\psi_{x} )]\nn
\eea
with $a_\m=a e_\m$, $e_0=(1,0), e_1=(0,1)$,
$\g^\pm_\m=\g_\m\mp r$ 
$F_{\m,\n}=d_\n A_\m-d_\m A_\n$ and 
$d_\n A_\m=a^{-1}(A_{\m,x+e_\n a}-A_{\m,x})$, 
%$:e^{\pm i e a A_\m(x)
%}:=e^{\pm i e a  A_\m(x)
% } e^{{1\over 2}
%e ^2  a^2  g^A_{\m,\m} (0,0)) }$ is the wick ordering (note that $a^2  %g^A_{\m,\m} (0,0)\le C$)  
$\tilde m=(m +4 r/a)$ and $r=1$ is the Wilson term. Note that if $1/a$ and $L$ are finite the integral is finite dimensional. 

We generalize the model adding a term $(1-\x) a^2 \sum_x \sum_\m (d_\m A_\m)^2$, $\x\le 1$ so that the bosonic action is given by ${1\over 2} a^2 \sum_x( \sum_{\m,\n} (d_\m A_\n)^2+\x \sum_\m (d_\m A_\m)^2)$. The original model is recovered with $\x=1$.

The correlations can be written as derivatives of the
generating function, 
\bea
&&e^{W_\x(J,B,\phi)}=\\
&&\int P(d\psi) P(d A) e^{-V(A+J,\psi)+(\psi,\phi)+a^2\sum_x B_x O}\label{ss}\nn
\eea
with $O=O(A+J,\psi)$ an observable, 
%is such that $O(A+J,\psi)=O(A+J+\partial \a,\psi e^{i \a})$. %As an example of $O$ we can consider the chiral current, which can be chosen %as $A^5_\m\bar\psi\g_\m\g_5\psi$.
and $P(dA)$
the gaussian measure 
with covariance
\be
\hat g^A_{\m,\n}(k)=
{1\over |\s|^2+M^2}(\d_{\m,\n}+ {\x \bar\s_\m \s_\n\over (1-\x) |\s|^2+M^2})
\ee
with $\s_\m(k)=(e^{i k_\m a}-1)a^{-1}$. 
%note that 
%For $\x<1$  $\hat g^A_{\m,\n}(k)$
%decays with $z=2$ while for $\x=1$ one has $z=0$. 
%This implies that for $\x=1$ $|g^A|_1\le C M^{-2}$ and $|g^A|_1\le C |\log a|$
%while for $\x=0$ $|g^A|_1\le C |\log a|$ and $|g^A|_\io\le C a^{-2}$.
%Note that for $\x<1$ then $|g^A_{\m,\n}|\le C/M^2$.

$P(d\psi)$ is the fermionic integration with propagator
$\hat g^\psi(k)= Z_\psi^{-1}   (\tilde k_\m \g_\m+a^{-1} m(k) I)^{-1}$ with $\tilde k_\m={\sin (k_\m a)\over a} $
and $m(k)=m+r a^{-1} (\cos a k_0+\cos a k_1-2)$; finally
%\be
%\begin{pmatrix}
%&{1\over a}(i\sin a k_0+\sin a k_1)
%&m+c(k)\\ &m+c(k)&{1\over a}(i\sin a k_0-\sin a k_1)
%\end{pmatrix}
%\ee
%
%
\be
V(A,\psi)=
a^2\sum_{x}  [O^+_{\m,x}G_{\m,x}^+(A) +O^-_{\m,x}G_{\m,x}^-(A)]
\ee
with $O^+_\m=Z_\psi\bar\psi_x (\g_\m-r) 
\psi_{x+a_\m}$ and
\be 
O^-_\m =-Z_\psi\bar\psi_{x+a_\m} (\g_\m+r)
\psi_{x}\ee  
and $G_\m^\pm =a^{-1} 
(e^{\pm i e a A_{\m,x}}-1)$. 
%If the observable $O$ is such that
%$O(A,\psi)=O(A+\partial \a,\psi e^{i \a})$ (called invariant observables)
%we define
%$\G_{\m_1,..,\m_n,\n_1,..\n_m}$, defined as the derivatives of $W_\x$ with %respect to $J_{\m_1,x_1},..,B_{\n_1,y_1}$.

If $M=0$ the model \pref{llll} invariant under the gauge transformation 
$A_{\m,x}\to A_{\m,x}+d_{\m}\a_{x}$ and $\psi_x\to \psi e^{-i e \a_x}$; if
$M\not=0$ the invariance is lost. 

\section{Ward Identities and $\x$-independence} 

If we restrict to observables  such that
$O(A,\psi)=O(A+d \a,\psi e^{-i e \a})$ (which we call invariant observables)
there is {\it gauge invariance in the external fields} also for $M\not=0$, that is 
\be
W_{\x}(J+d \a , e^{-i e \a}
\phi,B)=W_{\x}(J, 
\phi,B)
\label{hh}\ee
This follows by performing in \pref{ss} the change of variables $\psi_x\to \psi e^{i e \a_x}$, with 
Jacobian equal to $1$ (the integral is finite-dimensional) and noting that $(e^{i e \a} \psi,\phi)=(\psi,\phi e^{-i e \a})$ and
\be
S_\psi(A+J,\psi e^{i e \a})=S_\psi(A+J+d\a,\psi)
\ee
\pref{hh} implies that
$\partial_\a W_{\x}(J+d \a , e^{-i e \a}
\phi,B)=0$.
We define
$\G_{\m_1,..,\m_n,\n_1,..\n_m}$ as the derivatives of $W_\x$ with respect to $J_{\m_1,x_1},..,B_{\n_n,x_n}$. 
%We can imagine that the boson mass is generated by some Higgs mechanism %so that only gauge invariant obersvables are meaningful.
By performing in \pref{hh} derivatives with respect to $\a$ and the external fields
we get the Ward Identities (expressing current conservation)
\be
\sum_{\m_1} \s_{\m_1}(p_1) \hat\G_{\m_1,..,\n_n}(p_1,..,p_{n-1} )=0\label{ss22a}
\ee
and 
\be
\s_\m(p) \hat\G_\m(p.k)=\hat S(k)-\hat S(k+p)\label{hh22}
\ee
where $\hat\G_\m(p,k)={\partial^3 W\over \partial \hat J_{\m,p} \partial \hat \phi_{k}  \partial \bar \phi_{k-p}}|_0$
is the vertex function and $\hat S(k)={\partial^2 W\over \partial \hat \phi_{k}  \partial \bar \phi_{k}}|_0$ is the 2-point function.

The conservation of current expressed by the above WI implies that
for invariant observables 
\be
\partial_\x W_{\x}(J, 0,B)=0\label{inf}
\ee
that is the averages are $\x$ independent. This follows from $\partial_\x \int P(dA) \int \prod_x d\psi_x d\bar \psi_x O=0$, with $O(A,\psi)$ invariant;
indeed
\bea
&&\partial_\x \int P(dA)\int \prod d\psi_x d\bar \psi_x O=\label{inf11}  \\
&&{1\over L^2 }\sum_p 
\partial_\x \hat g^{-1}_{\m,\n}(p) \int P(dA) A_{\m,p} A_{\n,-p}
\int \prod d\psi_x d\bar \psi_x O\nn
\eea
from which we get, using that $A_{\m,p}=\hat g^A_{\m,\r} {\partial\over \partial A_{\r,-p}}$
\bea
&&\hat g^A_{\m,\r' }(p) \partial_\x (\hat g^A(p))^{-1}_{\m,\n}   \hat g^A_{\n,\r }(p)\\
&& 
{\partial^2\over \partial \hat J_{\r,p}\partial \hat J_{\r',-p}} \int P(dA) \int \prod_x d\psi_x d\bar \psi_x O(A+J,\psi)|_0 \label{ss11}\nn
\eea
By noting that
\be \partial (\hat g^A)^{-1}=- (\hat g^A)^{-1} \partial_\x \hat g^A( \hat g^A)^{-1}\ee 
and   $\partial_\x \hat g^A$ is proportional to $\bar\s_\m \s_\n$, 
by using \be \partial_\a \int P(dA) \int \prod_x d\psi_x d\bar \psi_x O(A+d \a,\psi)|_0=0\ee
then \pref{inf11} is vanishing. 

\pref{inf} ensures that the averages does not depend on $\x$,
so that one can set $\x=0$ in the boson propagator, that is the non decaying part
of the propagator does not contribute. In perturbation theory
the scaling dimension with $\x=0$ ($z=2$) and $\x=1$ ($z=0$)
is, if $n$ is the order, $n_A$ the number of $A$ fields and $n_\psi$ the number of $\psi$ fields
\be
d+(d-z-2)n/2-(d-1)n_\psi/2-(d-z)n_A/2
\ee hence in $d=2$ the theory is dimensionally renormalizable with $\x=1$ and superrenormalizable with $\x=0$ (in $d=4$
one pass from non-renormalizability to renormalizability). The lattice regularization ensures that i
the theory remains perturbatively superrenormalizable, as with dimensional regularization.
We will investigate the validity of this property at a non-perturbative level.

%Note that with momentum regularization, as used in exact solutions
%for the continuum case, both \pref{ss22a} and
%\pref{inf} are not true. 

Finally, we define the
axial current as $j_\m^5=Z^5 \bar\psi_x\g_\m\g_5\psi_x$, where $Z^5$ is a constant to be chosen so that the electric charge of the chiral and e.m. current are the same, 
defined as the amputated part of the 3-point correlation at zero momenta (see \cite{AB1}),
that is 
\be
\lim_ {k,p\to 0} {\partial^3 W\over \partial \hat B^5_{\m,p}\partial\hat \phi_k \partial \bar \phi_{k-p} }|_0/ 
{\partial^3  W\over \partial \hat J_{\m,p}\partial\hat \phi_k \partial \bar \phi_{k-p} }|_0=1\label{lap} \ee
where the source term is $(B_\m^5, j_\m^5)$.
The axial current is non conserved even for $m=0$, due to the presence of Wilson term, and one has
\be
\s_{\m}(p) \hat\G^5_{\m,\n}(p)=H_\n(p)\ee
with  $\G^5_{\m,\n}$ the derivative of $W$ with respect to $B_{\m,x_1}, J_{\n,x_2}$. $H_\n$ is called the anomaly and in the non-interacting case
$V=0$ one gets if $m=0$ $H_\m={1\over 2\pi}\e_{\m,\n}p_\n+O(a p^2)$
(lattice or dimensional regularization \cite{GRb} produce the same result) and $Z^5=1$. In the interacting case $H_\n(p)$
is a series in $e$ and the non renormalization property means that all higher orders corrections vanishes. 
%In contrast with a momentum regularization 
%in the non interacting $V=0$ case one gets $p_\m <j_\m^5 j_\n>=
%{1\over 4\pi}\e_{\m,\n}p_\n$ and   $p_\n  <j_\m^5 j_\n>={1\over 4\pi}%%%\e_{\m,\n}p_\n$, $Z^5=1$, that is either current and axial current are not %conserved. The momentum regularization is the one used in the continuum
%exact solutions.

\section{Integration of the boson fields} 

We can integrate the boson field
\be
\int  P(dA) e^{-V}=e^{\sum_{n=0}^\io{(-1)^n\over n!}\EE^T_A(V;n)}\equiv e^{V^N(\psi,J)}
\ee
where $\EE^T_A$ is the truncated expectation, that is the sum of connected diagrams, and 
$V^N=$
\bea
&&a^2 \sum_{x}\sum_\e  a^{-1} e^{-{1\over 2}
e^2  a^2  g^A_{\m,\m} (x,x)} e^{i a e J_{\m,x}}  O^\e_\m+\label{ap1}\\\
&&\sum_{n,m} a^{n+m} \sum_{\underline x,\underline y}\sum_{\underline\m,\underline \e}
[\prod_{j=1}^n
 O^{\e_j}_{\m_j,x_j}] [\prod_{k=1}^m G^{\e_j}_{\m_j,y_j}(J)] W_{n,m}(\underline x, \underline y)\nn
\eea
Note that $a^2  g^A_{\m,\m} (x,x)\le C$.
We call $a=\g^{-N}$, where $\g>1$
is a scaling parameter.
\newtheorem{thm}{Theorem}
\begin{thm}
The kernels in \pref{ap1} for $n\ge 2$ verify, 
\be |W_{n,m} |\le C^{n+m} e^{2(n-1)}   \g^{N(2-n-m)}
(|g^A|_1)^n
%with $\tilde C_0=M^{-2(n+m-1)}$
\label{ass}
\ee
\end{thm}
\noindent
{\it Proof.} A convenient representation for $\EE^T_A$ is given by the following formula \cite{Br}
\be
\EE^T_A(\prod_{k=1}^n e^{i \e_k  a  A_{\m_k,x_k}}) 
=\sum_{T\in  \TT }\prod_{i,j\in T} V_{i,j}
\int dp_T(s) e^{-V_T(s)}\label{for1}
\ee 
where $V_{i,j}=e^2 a^2 \EE_A(A_{\m_i,x_i}  A_{\m_j,x_j}) $,
$\TT$ is the set of tree graphs $T$ on $X=(1..,n)$, $s\in (0,1)$ is an interpolation parameter, $V_T(s)$ is a convex linear combination of $V(Y)=\sum_{i,j\in Y}\e_i \e_j V_{i,j}$, $Y$ subsets of $X$ and $dp_T$ is a probability measure. 
The crucial point is that
$V(Y)$ is stable, that is 
\be
V(Y)=\sum_{i,j\in Y} V_{i,j}=
a^2 e^2 \EE_A( [\sum_{i\in Y}  \e_i  A_{\m_i,x_i}]^2)\ge 0
\ee 
Therefore one can bound the exponential $e^{-V_T(s)}\le 1$ finding
\bea 
&&|W_{n,0}|\le 
 C^n a^{-n} {1\over n!}
 \sum_{T\in  \TT } \prod_{(i,j)\in T} a^2 e^2  |g^A(x_i,x_j)|_1\le \\
&&\sum_{T\in  \TT } 
C^n e^{2(n-1)} a^{n-2}  (|g^A|_1)^n\le  C^n e^{2(n-1)}  \g^{N(2-n)}
(|g^A|_1)^n\nn
%M^{-2(n-1)}\nn
\eea
%
%where we have used that $|g_A|_1=\int dx |g_A|\le C M^{-2}$; 
With 
$m\not=0$ we get an extra $a^{-N m}$, so that one recovers the dimensional factor $\g^{N(2-l/2-m)}$.
\qed
%The above bound says that the generating function is well defined for $L,1/a$ finite and $\x\le 1$, as it has been reduces to a finite dimensional Grassmann integral %($|v|_1$ is finite for $\x\le 1$). 
%If we restrict to gauge invariant observables by \pref{ass} we can do the convenient choice $\x=0$, where
%$|g_A|_1=\int dx |g_A|\le C M^{-2}$ indipendently on $a$. If we imagine that
%the boson mass is generated by some Higgs mechanism, then only gauge invariant observables could be indeed considered.

For $\x=0$ $|g^A|_1\le C M^{-2}$ and $|g^A|_\io\le C |\log a|$
while for $\x=1$ $|g^A|_1\le C |\log a|$ and $|g^A|_\io\le C a^{-2}$.
We write $W_{n,m} =\l^{n-1} \bar W_n$ with $\l=e^2$ and $\bar W_n$ bounded.
The normalization $Z_\x$ in the analogue of \pref{llll} is intere and $\x$ independent;
our strategy is to prove that
$Z_0=1+O(\l)$ and is analytic together with correlations for $|\l|\le \l_0$ with $\l_0$ $a,L$ independent;
the same therefore is true for $Z_1$, and as the numerator \pref{llll} is intere, than \pref{llll} is analytic in $|\l|\le \l_0$ and equal to the $\x=0$ case.
It remains then to prove that the correlations with $\x=0$ are analytic for  $|\l|\le \l_0$  and $Z_0=1+O(\l)$.

The factor $D=2-n-m$ is the scaling dimension, and the terms with $D<0$ are irrelevant.  The marginal term for $\x=0$ is 
$\EE^T_A( V;2)=$
\be \sum_{\e_1,\e_2}   a^4 \sum_{x_1,x_2} e^{i \e_1 a J_{\m,x_1}}
O^{\e_1}_{\m,x_1}e^{i \e_2 a J_{\m,x_2}}
O^{\e_2}_{\m,x_2} \l v_{\m,\e_1,\e_2}
\ee
where $\l v_{\m,\e_1,\e_2}=\EE^T_A( e^{i \e_1 e a A_{\m,x_1}};  e^{i \e_2 e a A_{\m,x_2}})$ and, $\l=e^2$
\bea
&&v_{\m,\e_1,\e_2}(x,y)=
e^{-{1\over 2}
e ^2  a^2  g^A_{\m,\m}(x_1,x_1) }\label{18}\\
&&e^{-{1\over 2}
e ^2  a^2  g^A_{\m,\m}(x_2,x_2) }(e^{-a^2 \e_1 \e_2 g^A_{\m,\m}
(x_1,x_2)}-1) e^{-2} a^{-2}\nn
\eea
which can be rewritten as
\be
\int_0^1 dt g^A_{\m,\m}(x_1,x_2)
 e^{-\tilde V(t)}
\ee
with 
\bea
&&2 \tilde V(t)=a^2  t \la (\e_1 A_\m(x_1)+ \e_2 A_\m(x_2))^2\ra \\
&&+a^2  (1-t) (g^A_{\m,\m}(x_1,x_1)+g^A_{\m,\m}(x_2,x_2))\nn
\eea
in agreement with \pref{for1}.
For definiteness we keep only the dimensionally non irrelevant terms considering
\be e^{W_1(J,B,\phi)}=
\int P(d\psi) e^{
\VV+G(B)+(\psi,\phi) }\label{ass}
\ee
with $G(B)$ is a generic source term for gauge invariant observables
and $\VV=$
\be
a^2 \sum_{x}\sum_\e  a^{-1} (e^{-{1\over 2}
e^2  a^2  g^A_{\m,\m} (x,x)} e^{i a e J_{\m,x}}-1)  O^\e_\m+\EE_A^T(V;2)\ee
Note that $a^2  g^A_{\m,\m}$
vanishes as $a\to 0$.
In the case of the chiral current \be G(B^5)=a^2 \sum_x Z^5 B^5_{x,\m}  \bar\psi_x\g_\m\g_5 \psi_x\ee

\section{Integration of the fermionic fields.} 

Our main result is the following

\begin{thm} For $|\l|\le \l_0 M^2$, 
with $\l_0$ independent on $a, m,M$ and $Z_\psi=1$
the correlations of 
\pref{ass} are analytic in $\l$; when the fermion mass is vanishing
the anomaly is $H_\m={\e_{\m\n}p_\n\over 2\pi }+O(a p^2)$. 
\end{thm}
%A naive estimate for $e_0$ would provide a value vanshing as $1/a$ or $L$ are
%sent to infinity; we prove that convergence holds in a finite domain.
%Non-perturbative and finite lattice effects effects do not spoil properties valid in perturbation theory, like
%the reduction of the degree of divergence, manifesting in the 
%the finiteness of wave function $Z_\psi=1$, and the anomaly non renormalization. 
%In the continuum with momentum regularization, both in exact 
%solutions \cite{GR} or Renormalization Group analysis \cite{M},\cite{F},
%one gets instead an infinite wave function renormalization and 
%an anomaly both in the vector and axial current,
%$p_\m <j_\m^5 j_\n>=\e_{\m,\n}p_\n(1-\t)\quad  p_\n  <j_\m^5 %j_\n>=\e_{\m,\n}p_\n(1+\t)$, $\t^e^2/4\pi$, as consequence of momentum %regularization.
\vskip.1cm

In order to integrate the fermionic fields
we introduce a decomposition of the propagator 
\be
g^\psi(x)=\sum_{h=-\io}^N g^{(h)}(x)\ee $\hat g^h(k)=f^h(k) \hat g(k)$
with $f^h(k)$ with support in $\g^{h-1}\le |k|\le \g^{h+1}$. 
One has to distinguish two regimes, the ultraviolet 
high energy scales $h\ge h_M$ with $h_M\sim \log M$ the mass scale, and the infrared regime $h\le h_M$. 
In the first regime, one uses the non locality of the quartic interation
\cite{Le},\cite{M},\cite{F}.
After the integration of the fields $\psi^N,
\psi^{N-1},..,\psi^h$, $h\ge h_M$ one gets an effective potential with kernels 
$W^h_{l,m}$ with $l$ fields ($l=2n$)
similar to \pref{ap1}, which can be written as an expansion in
$\l$ and in the kernels
$W^k_{2,0}$, $W^k_{4,0}$,$W^k_{2,1}$ with $k\ge h+1$.
Assuming that, for $k \ge h+1$ one has
$|W^k_{2,0}|_1\le   \g^h\l /M^2 $, $|W^k_{4,0}-v \l|_1
\le  \l^2/M^2$ 
and $|W^k_{2,0}-1|_1\le  \l^2/M^2$
then we get 
\be
|W^h_{l,m}|_1\le C^{l+m} (\l /M^2)^{d_{l,m}}
\g^{h(2-{l\over 2}-m)h}\label{ind1}
\ee
for $d_{l,m}=max(l/2-1,1)$ if $m=0$, and 
$d_{l,m}=max(l/2-1,0)$ if $m=1$. The proof of \pref{ind1}  is based on the analogous of formula 
\pref{for1} for Grassmann expectations
\be
\EE^T_\psi(\prod_{k=1}^n \tilde\psi(P_i)) 
=\sum_{T\in  \TT }\prod_{i,j\in T} V_{i,j}
\int dp_T(s) \det G\label{for}
\ee 
and the use of Gram bounds for get an estimate on $\det G$; in addition one uses that $|v|_1\le C M^{-2}$, 
$|g^h|_1\le C \g^{-h}$, $|g^h|_\io\le C \g^{h}$. 
We proceed by induction to prove the assumption. 
One needs to show that there is an improvement in the bounds due to the non locality of the boson propagator.
The kernel of the 2-point function
$
W_{2,0}^h(x,y)$ which can be written as sum over $n$ of truncated expectations and, if $\EE^T_{h,N}$
is the truncated expectation with respect to $P(d\psi^{[h,N]})$
\be
\l {\partial\over \partial\phi^+_x}{a^4\over (n-1)!} \sum_{x_1,x_2} v_{\m,\e_1,\e_2}
\EE^T_{h,N}({\partial\over \partial\phi^-_y}
O^{\e_1}_{\m,x_1}O^{\e_2}_{\m,x_2};V;...)\nn 
\ee
By using the property, if $\tilde\psi(P)=\prod_{f\in P} \psi_{x_f}$
\bea
&&\EE^T_{h,N}(\tilde\psi(P_1\cap P_2)...\tilde\psi(P_n))=\EE^T_{h,N}(\tilde\psi(P_1)\tilde\psi(P_2)...\tilde\psi(P_n))+\nn\\
&&\sum_{K_1,K_2\atop  K_1\cup K_2=3,..,n}
\EE^T_{h,N}(\tilde\psi(P_1)\prod_{j\in K_1}\tilde\psi(P_i))\EE^T_{h,N}(\tilde\psi(P_2)\prod_{j\in K_2}\tilde\psi(P_i))\nn
\eea
we get, omitting the $\e,\m$ dependence, $W_{2,0}^h(x,y)=$
\bea
&&\l a^4\sum_{z_1,z_2}  v(y,z_1) g^{[h,N]}(y+a_\m,z_2) W^h_{0,2}(z_2,x)
W_{1,0}(z_1)\label{pp}\nn\\
&&\l a^{-1} (e^{-{1\over 2}
e^2  a^2  g^A_{\m,\m} (0))}-1)  a^2 \sum_z g^{[h,N]}(x,z) W_{2,0}^h(z,y)\nn\\
&&+\l a^4 \sum_{z_1,z_2}
v(y,z_2) g^{[h,N]}(y+a_\m,z_1) W^h_{2,1}(z_2; z_1,x) 
\eea
The second term is bounded by \be C \l \g^h \g^{-h} a \log a\le \l/M^2  \g^h/2 \ee
for $a$ small enough.
The first term contains $\hat W_{1,0}(0)=0$.
%The action is invariat under $A_\m(x)\to -A_\m(-x-a_\m)$
%and  $\psi_x\to \psi_{-x}\s_0$; indeed 
%we get  $\sum \bar\psi_x \s_\m e^{i a e A_\m(x)} \psi_{x+a_\m}
%\to  \sum \bar\psi_{-x}  \s_0\s_\m e^{-i e e A_\m(-x-a_\m)} %%\s_0\psi_{-x-a_\m}$ so that $\sum_x W_{0,1}(x) =-\sum_x %W_{0,1}(-x-a_\m) =0$.
%which is equal to $-\sum \bar\psi_{x+a_\m} e^{-i A_\m(x)} \psi_{x}$.
%$-\sum  \bar\psi_{x+e_\m} \s_\m e^{-A_\m(x)}\psi_{x}$; moreover
% $\sum \bar\psi_x e^{A_\m(x)} \psi_{x+e_\m}\to 
%\sum \bar\psi_{-x}  e^{-A_\m(-x+e_\m)} \psi_{-x-e_\m}$
%which is equal to $\sum  \bar\psi_{x+e_\m} e^{-A_\m(x)}\psi_{x}$.
%This implies that $<J_\m(x)>=-<J_\m(x+e)>$ so that $\sum_x %<J_\m(x)>=0$.
Regarding the last term we get a bound
\be
\sup_{z_1,z_2}  |a^2\sum_{y}
v(y,z_2) g^{[h,N]}(y+a_\m,z_1)| a^2\sum_ {z_2,z_1} |W^h_{2,1}(z_2; z_1,0)| \label{ni}
\ee
\insertplot{220}{110}
{\ins{10pt}{70pt}{$W^h_{2,0}$}
\ins{50pt}{65pt}{$=$}
\ins{75pt}{90pt}{$W^h_{0,1}$}
\ins{100pt}{45pt}{$W^h_{2,0}$}
\ins{130pt}{65pt}{$W^h_{2,0}$}
\ins{180pt}{80pt}{$W^h_{2,1}$}
\ins{110pt}{65pt}{$+$}
\ins{160pt}{65pt}{$+$}
}
{figjsp467aa11}
{\label{h2} Graphical representation of \pref{pp}
} {0}
By using the inductive hypothesis $a^2\sum_ {z_2,z_1} |W^h_{2,1}(z_2; z_1,0)| \le C$ we get for \pref{ni} the bound
\bea
&&\l C_1 |a^2\sum_{y}
|v  g^{[h,N] }|\le \l C_ 1 [a^2\sum_{y}
|v|^3]^{1\over 3}\times\label{y}\\
 &&
[a^2 \sum_y  (g^{[h,N]})^{3\over 2}]^{2\over 3}\le \l M^{-2} C_2 \g^h \g^{-{4\over 3}(h-M)} \le \l M^{-2} \g^h/2\nn
\eea
for $h\ge h_M$, for $h_M=C \log M$ and $C$ large enough. Note that the above estimates uses
crucially that $\x=0$; for $\x=1$ $[a^2\sum_{y}
|v|^3]^{1\over 3} $ would be non bounded uniformly in $N$.

A similar computation can be repeated for  $W_{2,1}$;
in particular for the quartic term one uses that the boubble graph is finite
$A=\int dk {\rm Tr} g(k) \g_\m g(k)\g_\n$ so that $|W_{1,2}|_1\le C\l/M^2 (\g^{h-M}+A |W_{1,2}|_1)$.
The above estimates 
work for $h\ge h_M$ and it says that the theory is superenormalizable up to that scale.

In the infrared regime
$h_m\le h\le h_M$, where $h_m=\log_\g m$ is the fermion mass scale, 
the multiscale integration procedure is the same as in the Thirring model with a finite cut-off \cite{BFM}. 
The theory is renormalizable in this regime and there is wave function renormalization at each scale
$Z_h\sim \g^{-\h h}$, $\h=O(\l^2)>0$ and an effective coupling with asymptotically vanishing function. The expansions converge therefore uniformly in 
$a, L,M$ and the limit $a\to 0, L\to \io$ can be taken.
\insertplot{245}{65}
{\ins{125pt}{35pt}{$+$}
\ins{55pt}{35pt}{$=$}
\ins{-14pt}{35pt}{$p_\m$}
\ins{-45pt}{35pt}{$(1\pm \t)$}
}
{figjsp467aa111}
{\label{h2} Graphical representation of \pref{spp1}
} {0}

\section{Anomaly non-renormalization} 

The average of the chiral current $\G^5_{\m,\n}={\partial^2 W\over \partial B_\m \partial 
J_\n}|_0$ for $m=0$ is expressed by a series in $\l$. 
It is convenient to introduce a continuum relativistic model 
$
e^{\tilde W(J,B,\phi)}=$
\be\int P_{\tilde Z}(d\psi)
 e^{-V+\tilde Z^+ (J,j)+\tilde Z^- (B,j^5)+(\psi,\phi)}\label{eff}
\ee
where $P_{\tilde Z}(d\psi)$ has propagator
${1\over \tilde Z}{\chi(k)\over \g_\m k_\m}$, with $\chi$ a momentum cut-off selecting momenta $\le \g^{\tilde N}$, and 
\be
V=\tilde Z^2 \tilde\l
\int dx dy v(x,y) j_{\m,x} j_{\m,y}\ee with $v$ exponentially decaying with rate rate $M^{-1}$ with quartic coupling $\tilde\l$; finally $j^+_{\m,x}
\equiv j_{\m,x}=\bar\psi_x\g_\m\psi_x$ and $j^-=\bar\psi_x\g_\m\g_5\psi_x$. 

The infrared scales $h\le h_M$ of the two models differs by irrelevant terms
and one can choose $\tilde\l$ 
and $\tilde Z, \tilde Z^-,\tilde Z^+$ as function of $\l$
so that the corresponding running couplings flow to the same fixed point for $h\to-\io$. As a result, defining 
\be
\tilde\G^5_{\m,\n}={\partial^2 \tilde W\over \partial B_\m \partial 
J_\n}|_0\ee
we get
\be
\hat\G^5_{\m,\n}(p)=Z_5 \tilde \G^5_{\m,\n}(p)
+R_{\m,\n}(p)
\ee
where $R_{\m,\n}(p)$ is a continuous function at $p=0$, while $\tilde\G^5_{\m,\n}(p)$ is not; this provide a relation between the lattice and the continuum model. 

The model \pref{eff}  has two global symmetries, that is $\psi\to e^{i \a} \psi$
and  $\psi\to e^{i \a \g^5} \psi$, but the WI acquires extra terms associated with the momentum regularization \cite{M}. 
In particular, if $\t=\tilde\l \hat v(0)/4\pi$, in the limit of removed cut-off $\tilde N\to\io$
\be
(1\mp \t) p_\m \tilde \G^\pm_{ \m}(k,p)={\tilde Z^\pm\over \tilde Z}
\g^\pm (\tilde S(k)-\tilde S(k+p))\label{spp1} \ee
where $\tilde\G^\pm_\m$ is the vertex function of 
are the vertex correlations of \pref{eff} of the current $(+)$ and chiral current $(-)$ and $\g^+=I$, $\g^-=\g_5$. In the same way the WI for the current is
\be
 p_\m \tilde \G^5_{\m,\n}={\tilde Z^+ \tilde Z^-\over 4\pi \tilde Z^2}
{\e_{\m\n}p_\m\over (1+\t)  }\quad  
p_\n \tilde \G^5_{\m,\n}={\tilde Z^+ \tilde Z^-\over 4 \pi \tilde Z^2}
{\e_{\n\m}p_\n\over  (1-\t)  }\label{ss1} \ee
By comparing \pref{spp1} with the Ward Identity \pref{hh22},
and using that the vertex and the 2-point correlations of lattice and continuum model coincide up to subleading term in the momentum, we get a
relation between the parameters $\t, \tilde Z^+, \tilde Z$
\be { \tilde Z^+ \over \tilde Z(1-\t)}=1	\ee 
Moreover the condition on $Z^5$ \pref{lap} and \pref{spp1} imply 
\be
{ \tilde Z^+ \over \tilde Z(1-\t)}=Z_5 { \tilde Z^- \over \tilde Z(1+\t)}=1\ee
from which $Z_5=(1+\t){ \tilde Z\over \tilde Z^-}$. By the Ward Identity \pref{ss22a} we get 
\be
p_\n \hat\G^5_{\m,\n}(p)={\tilde Z^+ \tilde Z^-\over 2 \pi \tilde Z^2}
{\e_{\n\m}p_\n\over  (1-\t)  }
+p_\n R_{\m,\n}(p)=0\label{fon2}
\ee
so that \be R_{\m,\n}(0)=- {\tilde Z^+ \tilde Z^-\over 2 \pi Z^2}
{\e_{\m\n}\over  (1-\t)  }=-(1+\t) \e_{\n\m}/Z_5\ee
Finally
\bea
&&p_\m\hat\G^5_{\m,\n}(p)=Z_5 p_\m [\tilde\G^5_{\m,\n}(p)+R_{\m,\n}(p)]=\nn\\
&&[(1-\t)\e_{\m,\n}-(1+\t)\e_{\n,\m}] p_\m/4\pi=1/2\pi\e_{\m,\n}p_\n
\eea
that is all the dependence of the coupling disappears.   
\qed
%The regularization implicit in the exact solutions \cite{GR} is 
%a momentum regularization and by \prefl{ss} the current is not conserved;
%therefre the $k_\m k_\n$ piece of the propagator contrubute and there is no %decreasing of divergence.

\section{Conclusions} 

We have analyzed a lattice version of the Sommerfield model.
Both the reduction of the degree of ultraviolet divergence, manifesting in the finiteness of the field renormalization, and
the Adler-Bardeen theorem hold at a non-perturbative level, in contrast with exact solutions in the continuum.
Non perturbative violation of perturbative results are therefore excluded. This provides support to the possibility
of a rigorous lattice formulation of the electroweak sector of the Standard Model with step exponentially small in the inverse coupling,
which requires an analogous reduction of degree of divergence. New problems include the fact that 
a multiscale analysis is necessary also for the boson sector, and
the 
fact that the symmetry is chiral and
anomaly cancellation is required; Adler-Bardeen theorem on a lattice 
is exact for non chiral theories \cite{GMP} but has subdominant corrections for chiral ones \cite{M1b}.

{\it Aknowledments}
This material is based upon work supported by a grant from the Institute
for Advanced Study School of Mathematics. I thank also GNFM and MUR through PRIN MAQUMA.

\end{document}